# Passive Measurement of Autonomic Arousal in Real-World Settings

Samy Abdel-Ghaffar, Isaac Galatzer-Levy, Conor Heneghan, *Senior Member, IEEE*, Xin Liu, *Member, IEEE*, Sarah Kernasovskiy, Brennan Garrett, Andrew Barakat, Daniel McDuff, *Member, IEEE*

*Abstract*—The autonomic nervous system (ANS) is activated during stress, which can have negative effects on cardiovascular health, sleep, the immune system, and mental health. While there are ways to quantify ANS activity in laboratories, there is a paucity of methods that have been validated in real-world contexts. We present the Fitbit Body Response Algorithm, an approach to continuous remote measurement of ANS activation through widely available remote wrist-based sensors. The design was validated via two experiments, a Trier Social Stress Test (n = 45) and ecological momentary assessments (EMA) of perceived stress (n=87), providing both controlled and ecologically valid test data. Model performance predicting perceived stress when using all available sensor modalities was consistent with expectations (accuracy=0.85) and outperformed models with access to only a subset of the signals. We discuss and address challenges to sensing that arise in real world settings that do not present in conventional lab environments.

¹This work was funded by Google, LLC. All the authors were employed by Google and received stock as part of their normal compensation.
*(Corresponding author: Samy Abdel-Ghaffar)*

Samy Abdel-Ghaffar is with Google Research, Mountain View, CA 94043 USA (e-mail: sabdel@google.com).
Isaac Galatzer-Levy is with Google Research, Mountain View, CA 94043 USA (e-mail: isaacgl@google.com).
Conor Heneghan is with Google Research, Mountain View, CA 94043 USA (e-mail: conorheneghan@google.com).
Xin Liu is with Google Research, Mountain View, CA 94043 USA (e-mail: xliucs@google.com).
Sarah Kernasovskiy was with Google Research, Mountain View, CA 94043 USA. She is currently unaffiliated.
Brennan Garrett is with Google Research, Mountain View, CA 94043 USA (e-mail: bdgarr@google.com).
Andrew Barakat is with Google Research, Mountain View, CA 94043 USA (e-mail: andrewbarakat@google.com).
Daniel McDuff is with Google Research, Mountain View, CA 94043 USA (e-mail: dmcduff@google.com).
Color versions of one or more of the figures in this article are available online at http://ieeexplore.ieee.org

## I. Introduction

The human stress response is a complex interplay of physiological and psychological processes that are critical for adaptation and survival [1], [2]. The ability to measure biological responses to stress during everyday life can open new pathways for learning and proactive interventions to address a wide range of physical and mental health risks. However, the ability to remotely index stress in real-world contexts remains elusive due to challenges in accessing stress physiology, defining measurement paradigms, and disambiguating it from other biological processes and responses to environmental stimuli.

Immediate biological responses to stress are commonly indexed using measures of autonomic arousal, a key component of the stress response that reflects the activity of the sympathetic and parasympathetic nervous systems, which regulate a wide range of bodily functions, including heart rate, breathing, and digestion [3]. Variability in autonomic arousal has been linked to a variety of psychological and physical health outcomes across the life course [4], [5].

Experimental studies using validated paradigms, such as the Trier Social Stress Test (TSST), have provided valuable insights into the mechanisms underlying the link between autonomic arousal and stress-related outcomes [6]. These studies have demonstrated that autonomic arousal is associated with changes in hormones, brain structures and functions, and emotion regulation, all of which play a role in long-term risk and resilience to stress.

Measuring autonomic arousal in real-world settings has the potential to provide important information about individuals' stress responses and their associated health risks [6], [7], [8]. However, remote measurement of stress is challenging as it relies on physiological metrics that are activated in response to myriad internal processes and external events that are not attributable to stress. Consistent with latent variable measurement theory [9], we hypothesize that each measure of physiology is a fallible indicator or weak signal of stress that, when combined with other related signals, will produce a more stable and accurate metric of the underlying latent construct of stress.

The field of affective computing has contributed substantially to research into stress sensing using wearable devices. Many physiological signals have been considered [10] with the most unobtrusive and scalable physiological measures for sensing stress including heart rate and heart rate variability from photoplethysmography. Wrist-worn electrodermal sensors were pioneered in the past 20 years and provide a more direct measure of sympathetic arousal [11].

These sensors have been used in many contexts from measuring stress in the workplace [12], to studying sleep [13]. However, despite this compelling research showing the utility in everyday life, there are a paucity of devices targeted broadly at consumers that leverage the capabilities of this technology for detecting, and raising awareness of, physiological stress.

Recent advances in wearable sensor technology have led to the development of a variety of devices that can measure physiological signals relevant to autonomic arousal, including heart rate, peripheral body temperature, and electrodermal activity (EDA) integrated into smartwatches. However, real-world settings present unique challenges for measuring autonomic arousal. These environments are often characterized by high levels of noise and variability [14], necessitating the development of signal processing and analysis methods that can effectively differentiate between true autonomic arousal and background noise. Real-world settings also introduce confounding everyday living conditions such as exercise, water exposure or device loose wear which present unique challenges not encountered in lab-based studies.

In the current work, we address the need for validated methods to measure autonomic arousal in real-world settings through the use of remote sensing technology. We introduce the Fitbit Body Response algorithm, debuted on the Fitbit Sense 2 smartwatch device, which utilizes ensemble models of well-established physiological measures of autonomic activity. We further evaluate these models' capacity to measure significant changes in autonomic arousal due to psychological stress and describe techniques to manage noise and confounding conditions associated with real-world settings.

## II. METHOD

### A. Data Collection

#### 1. Participants

The current study protocol was approved by the Advarra Institutional Review Board (IRB# Pro00054030). Informed consent was collected from all participants before study onboarding and subjects were compensated via a $50USD Google store cash incentive. Forty-five healthy adult English speakers (n=20 male) participated in both the virtual lab stress induction and free-living study portions (see Table 1 for detailed demographics), and were well-balanced across sex (males: M[SD]=39.6[9.2] years; females: M[SD]=37.0[8.5] years). An additional forty-two (n=22 male) healthy adult English speakers participated in only the free living study portion and were also well-balanced across sex (males: M[SD]=38.6[8.7] years; females: M[SD]=37.1[6.8] years). Potential participants were Google employees that were located in the USA, at least 21 years old, capable of giving informed consent and willing to complete surveys, go about their daily activities while wearing Fitbit wearable devices, and to participate in stress induction activities. Participants were excluded if they were regular tobacco users or had any self-reported psychiatric conditions.

#### 2. Experimental Protocol

*i. Trier Social Stress Test (TSST)*

To induce a physiological stress response under controlled conditions, the Trier Social Stress Test (TSST) was used, which is considered the "gold standard" for lab-induced stress procedures. The TSST induces social-evaluative stress via a mock job-interview and a surprise arithmetic task, and reliably induces subjective stress and sympathetic arousal responses as measured via physiological signals. Due to COVID-related social distancing protocols and mandatory work-from-home policies in place at the time of the study, an online version of the protocol was adapted utilizing video-conferencing software for use in this study. Subsequent studies have validated the effectiveness of similar approaches [15]. For this study, two 20-minute interviewer-panel videos were created (one all male panel, one all female panel). The 3-member interview-panel maintained a flat affect and moved minimally during the entire 20 minute videos. For each subject, the opposite-sexed video was used during the instructions, mock-job interview and arithmetic task portions of the study.

Subjects were instructed to wear their Fitbit study devices for the hour prior to the study. Subjects joined a video-conferencing call with the experimenter, after which they were told to relax calmly at their desks for a 15-minute baseline period. Following this was a 10-minute anticipatory period consisting of 5 minutes of instructions for the mock job-interview and 5 minutes of interview preparation or note taking. Prior to administering the instructions, subjects were asked to join a break-out room from within the video conferencing software. This break-out room was deemed the "study room". The experimenter had already joined the "study room" using a second video-conferencing account, and was playing the opposite-sexed interviewer-panel video as the background of the "study room". From the subject's perspective, it appeared that there was a live video feed of the interview-panel, which was actually a pre-recorded video. The experimenter began recording the video call from this point onwards, and spoke the instructions to the subject from a different account. After instructions were administered, the subject left the "study room" and returned to the "safe room" where they completed a 5-point stress ecological momentary assessment (EMA) survey (see Stress Labels section for more) and Spielberger State-Trait Anxiety Inventory (STAI) before doing the interview preparation. After the interview preparation period, the subject's notes document was deleted and the subject was asked to enter the "study room" for the 10-minute stressor period consisting of a 5-minute mock-job interview and subsequent surprise 5-minute arithmetic task. During the stressor period, the experimenter's audio and video feeds were turned-off and they communicated with the subject via chat messages. During the mock-interview, if the subject was quiet for 20 seconds then the experimenter sent a message asking them to please continue. After the arithmetic task was complete, the subject was asked to return to the "safe room" where they were debriefed for 10-minutes. A 2nd stress EMA was administered, followed by a viewing of a calming video and a 2nd STAI survey. Subjects were given a debriefing document with an explanation of the purpose of the study.

Finally, subjects were asked to rest calmly again for a 15 minute recovery period followed by a 3rd stress EMA.

*ii. Free Living*

To quantify performance of the Body Response algorithm in an ecologically valid setting, Fitbit physiological data was collected while subjects engaged in their everyday life activities over the course of 1 week. Subjects were shipped a wrist-worn Fitbit prototype device before being onboarded to the study via a video-conferencing session. During onboarding, surveys were conducted and instructions for the free-living portion of the study were given. Subjects were instructed on device usage and shown how to enroll in the Fitbit Guided Program used to collect the stress log labels. During onboarding, subjects were asked to wear their device for 7 days, during all waking hours, and respond to the 5-7 daily mobile phone stress log notifications (see Stress Labels section for more). Stress log notifications were sent between 8am-10pm local time. Additionally, subjects were sent daily emails that identified periods with increasing EDA signal from the previous day and asked to label each one. The TSST was administered at the end of the free living week.

*B. Stress Labels*

*1. TSST*

Traditionally, time during the baseline and recovery periods are labeled as not-stress and time during the anticipation and stressor periods are labeled as stress. Labels for these traditional stress/no-stress periods were generated. However, inspection of the physiological time-series data showed evidence of changes consistent with previously reported autonomic arousal events during the baseline period, at varying offsets during both the anticipation and stress periods, and well into the recovery period. Because the primary aim of this study was to build an algorithm that can identify the onset of autonomic arousal events, and not to conduct a traditional TSST analysis, we manually created stress/no-stress labels that did not follow the traditional labeling approach. Minutes which showed changes in the physiological signals commensurate with an autonomic arousal event were labeled stress, and all other minutes labeled no-stress, independent of which of the 4 TSST periods they fell in. This resulted in 74 total stress events, where the majority of subjects had 1 or 2 autonomic arousal events, and a few had 3 or 4 events. A total of 1028/2513 TSST minutes had a stress label. As described in the results section, model performance of the traditional TSST labels was higher than for the manually created stress labels, but the features learned did not capture the dynamics of an autonomic stress response, rather baseline shifts apparent across the traditional stress and no-stress periods.

During the TSST, two different surveys were administered. An ecological momentary assessment (EMA) was administered on the 1-5 Likert scale as used in the stress logs (see Stress Logs section below) at 3 time points during the TSST: immediately following the instructions, immediately following the stress induction, and at the end of the recovery period. These EMAs were used to validate that perceived stress had been induced via the protocol. Additionally the Spielberger State-Trait Anxiety Inventory (STAI) survey was administered twice, alongside the first and second EMAs following instructions and stress induction.

*2. Stress Logs*

A Guided Program within the Fitbit mobile app was used to collect stress log data. This provided an user interface, notification system, and data synchronization services. Subjects were shown how to enroll in the guided program during onboarding. Once a subject enrolled, the Guided Program would deliver 5-7 stress logging notifications daily. Notifications were delivered a minimum of 45 minutes apart. A previously created HR/HRV-based stress detection algorithm was used to detect potential moments of physiological stress response. If the HR/HRV-based algorithm detected physiological stress, and it had been 45 minutes since the last notification was sent, then a new notification was sent. If 2.75 hours have passed since the last notification, then a randomly timed notification was sent between 0-20 minutes later, sampled uniformly. A minimum of 5 notifications and maximum of 7 notifications were sent each day.

The stress log notification asked subjects 2 questions. The first question utilized a likert-type scale asking "How stressed did you feel at HH:MM:SS AM/PM?" with responses ranging from 1 (not at all stressed) to 5 (extremely stressed). The second multiple-choice question asked "How long before HH:MM:SS AM/PM did you feel this level of stress (please use your best judgment)?" with time-range responses ranging from "0-5 minutes ago" to "60+ minutes ago". Stress/no-stress labels were created from these notifications by assigning values of 1 or 2 on the first question a no-stress label, and values 3, 4 or 5 a stress label. The duration of the label was calculated by taking the average of the 2 times specified in the answer to question 2 (e.g. for 0-5 minutes, 3 minutes was and for 60+ minutes, 60 minutes was used).

*3. Retrospective EDA Surveys*

In order to get labels for minutes with EDA responses, and since the stress log notifications were triggered via an HR/HRV based algorithm without the benefit of the EDA data stream, retrospective surveys were filled out by subjects. An automated script ran every night after subjects synced their EDA data. This script identified EDA events where the EDA signal showed a meaningful increase in SCL value. A survey was automatically created for each user which asked them to identify the cause of each of the EDA events with the following possible responses: Heat/Exertion, Humidity, Stress, or Unknown. Stress/no-stress labels were created using these retrospective EDA surveys. A stress label was assigned to all EDA events where users responded "Stress", and a no-stress label was assigned to all EDA events where users responded either "Heat/Exertion", "Humidity" or "Unknown". The duration of the stress/no-stress label was the same as the duration of the EDA event identified.

*C. Input Signals*

In the following sections we describe the mechanics of the signal processing pipeline and classification of stress. We will discuss the raw signals, preprocessing, confounder removal, imputation of missing data, normalization, feature extraction, classification and post-processing. Fig. 2 provides a high-level overview of the pipeline.

*1. Heart Rate (HR)*

A validated algorithm [16] was used to extract heart rate (HR) at 1 Hz frequency from the photoplethysmography (PPG) signal. Missing data due to lack of contact between the wrist and Fitbit device or excessive motion is linearly interpolated over 1 minute time blocks. Minutely HR data was calculated by taking the mean of the interpolated, secondly data across non-overlapping 1 minute windows.

*2. Heart Rate Variability (HRV)*

Nine standard HRV metrics [17] were calculated every minute over a sliding 5-minute window, namely: R-R mean, R-R median, R-R 20th percentile, R-R 80th percentile, R-R Shannon Entropy, R-R differences Shannon Entropy, SDNN, RMSSD, & PNN30. An on-device peak detection algorithm identified PPG-based R-wave peaks from which R-R intervals were calculated. R-R intervals are susceptible to noise from multiple sources, including movement, electronic noise, missed heartbeats, etc. To account for noise, outliers were removed from each sliding 5-minute window using the median-filter based approach outlined in Natarajan et al. [18]. The percentage of each 5-minute window with valid R-R intervals was calculated, and windows with less than 20% valid R-R intervals or a maximum gap larger than 10 seconds were deemed too noisy for use in the Body Response algorithm.

*3. Electrodermal Activity (EDA)*

The electrodermal activity (EDA) sensor is used to infer sympathetic arousal via changes in micro-sweat levels on the dorsal surface of the wrist. Two electrodes on the back of the device measure changes in conductance, which varies with skin moisture levels. EDA data is sampled at 200Hz, downsampled to 25hz via a boxcar filter, and smoothed with 5-minute median and low-pass filters [19]. Minutely tonic EDA slope and magnitude are then calculated. Due to the nature of sensing mode operation, EDA data is only on during non-exercise wake-periods.

*4. Skin Temperature (ST)*

Skin temperature (ST) is measured via a temperature sensor located near the wrist facing surface of the device. Measurements are sampled every 10 seconds. Minutely slope and magnitude values are calculated via simple linear regression for use in the Body Response algorithm. Skin temperature signals are available whenever EDA signals are available.

*5. Confounder Filtering*

Ensembles of physiological signals including those used in the Body Response algorithm are known to be sensitive to detecting acute stress [20], but not particularly specific. Several known user behaviors can result in signal changes which mimic autonomic arousal events. These include exercise, exposure of the EDA electrodes to water, and loosely wearing the device resulting in intermittent EDA electrode contact with the wrist. Filters were added to the Body Response algorithm to remove minutes where these confounding behaviors were inferred to be present.

An exercise detection algorithm developed in prior unpublished work was used to independently exclude minutes where EDA and HR were unusable. The algorithm uses a logistic regression classifier across aggregates of a 10-minute sliding window of multiple Fitbit 3-axis accelerometer metrics to infer a probability of exercise vs sedentary behavior per minute. The HR signal is deemed unusable when the exercise detection algorithm determines the current minute is an exercise with at least 80% probability. The EDA signal is deemed unusable when exercise is 90% probable or higher and the EDA signal of the current minute is significantly larger than a 10-minute sliding mean.

Any non-sweat moisture introduced between the wrist and EDA electrodes, such as from water during hand-washing or showers, will result in EDA signal increases which can manifest as false positive sympathetic arousal events. To identify and remove minutes confounded by water exposure, a filter was developed which locates minutes with high variation in the on-device barometric pressure sensor and EDA signal increases.

Wearing the Fitbit device loosely can result in intermittent contact between the EDA electrodes and the dorsal wrist surface. When contact is lost, conductance drops to open-circuit levels, then rebounds to true readings once contact is reestablished. This results in artificially generated EDA drops and spikes which can mimic sympathetic arousal events. To identify minutes when the EDA signal is confounded by loose wear, the percentage of each minute with open-circuit level EDA signal was calculated. Minutes were removed from the Body Response algorithm with no contact for 50% or more of the minute, plus 5 minutes post to allow for signal stabilization.

### III. ANALYSIS

*A. Feature Extraction & Selection*

Features for the Body Response algorithm were calculated by applying aggregation functions across sliding windows of input signals. The 14 input signals used here comprise minutely values for the 9 HRV metrics (see HRV section above), mean HR, EDA slope/magnitude, and skin temperature slope/magnitude. To account for inter-subject variation in physiology, the 14 input signals were normalized per-user via z-scoring. Per-user means and standard deviations were calculated from all available data from the 1-week free living data. Both the original and per-user normalized versions of the input signals were used for model training and evaluation, resulting in a total of 28 input signals. Windows with a length of 31 minutes (current minute + 30 minute window) and a stride of 1 minute were created, resulting in a 31-minute window for every study minute. Only windows containing valid data for the most recent minute and having at least 20% data availability were kept for model training and evaluation. Within each window, missing data between valid data points was linearly interpolated, and leading missing minutes were backfilled. A Python package called tsfresh (https://tsfresh.readthedocs.io/en/latest/text/list_of_features.html) was used to apply common time-series aggregation functions to input signal windows. Each aggregation function (e.g. linear trend) can output multiple metrics (e.g. slope, bias, correlation). A total of 46 aggregation functions (see Table 2 for list of aggregation functions selected in Body Response algorithm), resulting in 206 unique features, were applied to each input signal. Univariate feature selection was done based on FDR corrected p-values, via the Benjamini-Hochberg procedure. The top 20 features within each of the 4 input signal groups (HR, HRV, EDA, ST) were selected for model estimation.

*B. Model Training*

Machine learning models were trained on the TSST dataset to predict minutes of stress/no-stress. All TSST minutes with valid signals across all 4 input signal groups were selected for classifier comparison (n=2513 windows). Logistic regression, random forests, and linear support-vector machines (SVMs) were trained on the aggregated tsfresh features (see Feature Extraction & Selection). A 2D convolutional neural network (Conv2D) was trained on the 31-minute input signal windows, after normalization and imputation. Leave-One-Subject-Out (LOSO) cross-validation was used for hyperparameter tuning and classifier performance comparisons. Due to the imbalanced nature of the training dataset (40.9% stress), label class weighting and a balanced accuracy CV metric were used. In addition to the feature selection described above, an L0 penalization term was used to regularize the logistic regression models.

Model evaluation was performed as described below, and the logistic regression classifier was selected based on a combination of performance and model interpretability. Input signal availability differed across HR, HRV, EDA, & ST. EDA & ST were available most frequently, followed by HR, then HRV. Given the real-world application of the body-response algorithm, 3 different models were trained that used decreasing numbers of input signals to allow for maximal temporal coverage. The smallest used only EDA & ST features, a middle-sized model added HR features to this, and the full model added HRV features to that. The final Body Response algorithm includes a model selection step, where the largest model with available input signals is selected to generate the stress/no-stress prediction. Finally, given the temporal nature of autonomic arousal events, a post-processing step was done to remove spurious predictions. Stress predictions that are less than 3 consecutive minutes are discarded, and predicted stress events that are closer than 5 minutes are stitched together into a single stress event.

*C. Performance Evaluation*

Stress labels for the free living period came from either the stress logs or EDA surveys (see Stress Labels section above). These labels have inherent noise in them, as they do not necessarily capture the exact start or end times of autonomic arousal events given the sparse sampling of time done during labeling. To account for this noise, events detected as stress during free living conditions which overlapped with actual stress events within +/- 10 minutes were considered "correct". Thus, to calculate free living performance metrics, the vector of stress/no-stress predictions was adjusted before performance metrics were calculated. If one or more predicted stress events overlapped with an actual stress event, then the entire actual stress event was marked as correct and the predicted stress events were discarded. Following this adjustment, all performance metrics were calculated between the actual and adjusted stress minutes.

This adjustment has the potential to change chance level performance (e.g. 50% accuracy), and so a permutation test was done to estimate chance performance across all metrics and provide p-values of each metric's significance level. The following procedure was done 1000 times for each metric, and chance level estimates were calculated by taking the mean across the resulting null distributions, while p-values were calculated by counting the number of values in the resulting null distributions that were larger than the actual metric values, and dividing that number by 1000, the size of the distribution. For each subject-day, all detected stress responses were identified and randomly shuffled in time with respect to the labels, keeping the size of each stress response the same and not allowing for overlapping stress responses. All the metrics were then calculated between the resulting shuffled predicted stress labels, and the ground truth labels.

| | | Virtual Lab (TSST) + Free-Living | | Free-Living Only | |
|---|---|---|---|---|---|
| | | Female | Male | Female | Male |
| **Demographic** | **Number (n)** | 25 | 20 | 20 | 22 |
| | **Age (years)** | 37.0 (8.5) | 39.6 (9.2) | 37.1 (6.8) | 38.6 (8.7) |
| | **Height (inches)** | 63.4 (2.3) | 70.6 (2.1) | 64.0 (3.4) | 68.2 (3.6) |
| | **Weight (lbs)** | 144.3 (29.5) | 175.0 (25.6) | 152.6 (36.0) | 185.6 (32.5) |
| | **BMI** | 25.1 (4.4) | 24.7 (3.3) | 26.2 (5.9) | 28.1 (4.6) |

**Table 1.** Demographics of the subjects in our study. A total of 87 subjects participated in the study, 45 in the virtual lab (TSST) portion and all 87 in the free-living portion.

| **tsfresh** Function | Description | cEDA+ Temp | cEDA+ Temp+ HR | cEDA+ Temp+ HR+ HRV |
|---|---|---|---|---|
| agg_linear_trend | Calculates a linear least-squares regression for values of the time series that were aggregated over chunks versus the sequence from 0 up to the number of chunks minus one. | ✓ | ✓ | ✓ |
| change_quantiles | First fixes a corridor given by the quantiles ql and qh of the distribution of x. Then calculates the average, absolute value of consecutive changes of the series x inside this corridor. | | | ✓ |
| energey_ratio_by_chunks | Calculates the sum of squares of chunk i out of N chunks expressed as a ratio with the sum of squares over the whole series. | | ✓ | ✓ |
| index_mass_quantile | Calculates the relative index i of time series x where q% of the mass of x lies left of i. For example for q = 50% this feature calculator will return the mass center of the time series. | ✓ | ✓ | ✓ |
| last_location_of_minimum | Returns the relative last location of the minimum value of x. The position is calculated relatively to the length of x. | ✓ | ✓ | ✓ |
| linear_trend | Calculate a linear least-squares regression for the values of the time series versus the sequence from 0 to length of the time series minus one. | ✓ | ✓ | |
| mean_change | Calculates the mean over the differences between subsequent time series values. | | | ✓ |
| median | Calculates the median of the values. | ✓ | ✓ | ✓ |
| number_cross_m | Calculates the number of crossings of x on m. A crossing is defined as two sequential values where the first value is lower than m and the next is greater, or vice-versa. If you set m to zero, you will get the number of zero crossings. | ✓ | ✓ | ✓ |
| number_peaks | Calculates the number of peaks of at least support n in the time series x. A peak of support n is defined as a subsequence of x where a value occurs, which is bigger than its n neighbours to the left and to the right. | | ✓ | ✓ |
| quantile | Calculates the q quantile of x. This is the value of x greater than q% of the ordered values from x. | ✓ | ✓ | ✓ |
| sum_values | Calculates the sum of all values. | ✓ | ✓ | |

**Table 2.** List of tsfresh aggregation functions used across 3 Logistic Regression Body Response models. See https://tsfresh.readthedocs.io/en/latest/api/tsfresh.feature_extraction.html for documentation of these functions.

## IV. Results

### A. Validation of Subjective Responses to Stress Events

Five-point ecological momentary assessments (EMAs) were taken at 3 points during the TSST (see Stress Labels within the methods section for more): immediately after the instructions, immediately after the stress induction, and at the end of the recovery period. Two-sided, paired t-tests between all 3 EMAs were significant at α<.05. The instructions EMA was less than the stress EMA (t(43)=-3.42, p=0.001, d=0.54) and was greater than the recovery EMA (t(40)=5.78, p<0.001, d=0.97). The stress EMA was greater than the recovery EMA (t(42)=9.73, p<.001, d=1.46). As expected, these findings show that the TSST successfully induced a subjective stress response across subjects.

### B. Physiological Changes During Stress Events

Fig. 1 shows the mean physiological input signals relative to all TSST stress response onsets. There are trends that HR increases, HRV drops, and EDA rises after the stress response begins. These distributions suggest that the input signals should be sensitive to changes in stress. However, there is also a large amount of variability in the individual signals suggesting that no one signal will be very specific on their own.

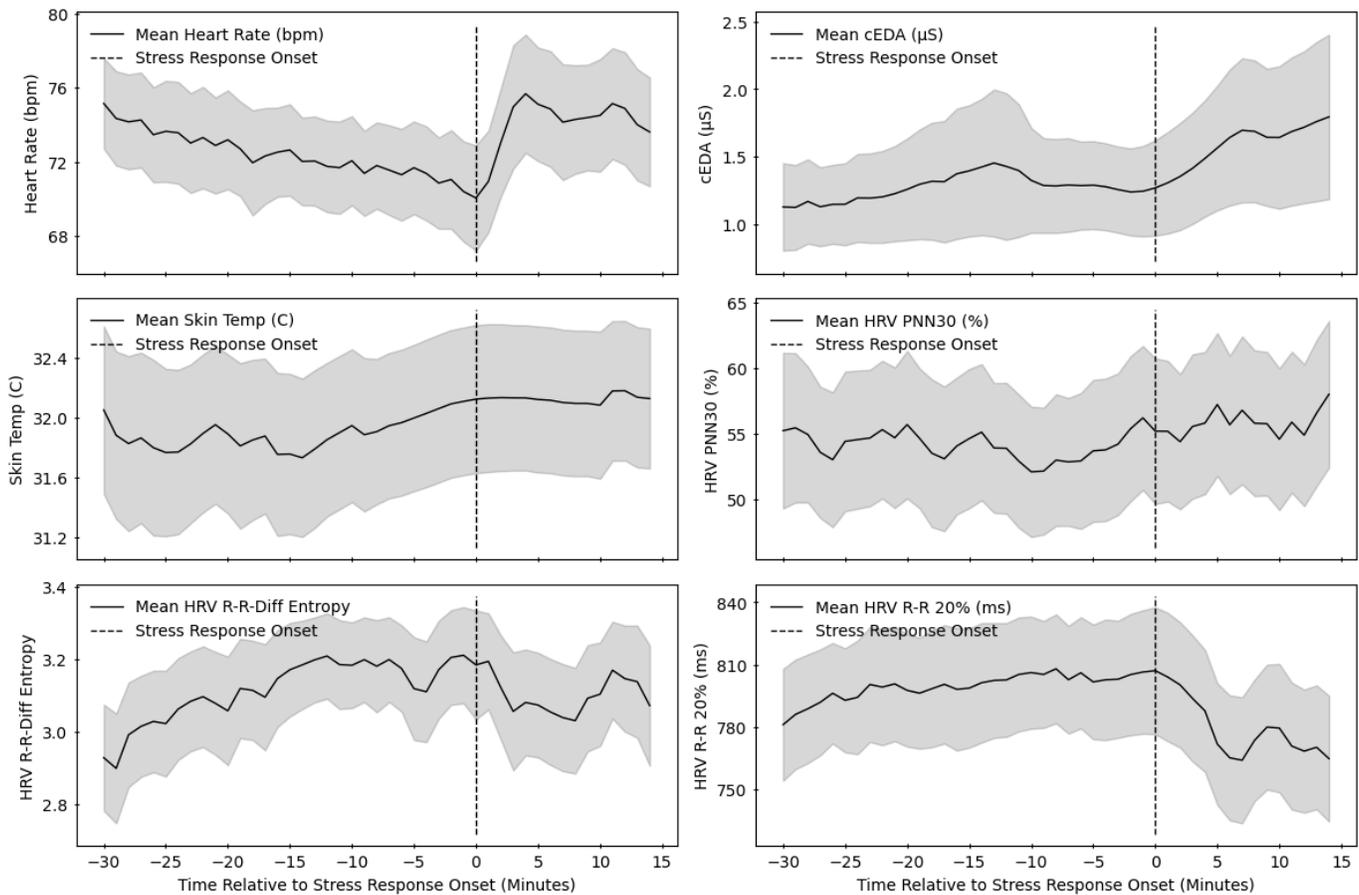

**Fig. 1.** Mean input signal time series relative to TSST stress response onsets across six physiological metrics of Heart Rate, Skin Temperature, EDA, & Heart Rate Variability.

### C. Trier Social Stress Test (TSST) evaluation

Four classification algorithms were used, in a leave-one-subject-out (LOSO) fashion, to train the virtual lab TSST data (see Table 3), namely random forest (RF), linear SVM (l-SVM), 2D convolutional neural network (Conv2D) and logistic regression (LR). The algorithms were selected as they cover a range of designs used in prior work and have a variety of desirable properties, some more interpretable (e.g., LR) and others are able to capture more complex non-linear relationships between features and labels. In order to compare algorithms clearly we decided to fix the operating point of each at a false positive rate of 29%, which was the FPR of the logistic regression classifier using an optimal probability threshold of 0.50. Accuracy across the four algorithms did not vary dramatically, with both linear models showing better balanced accuracy (l-SVM=0.73, LR=0.73) than the non-linear models (RF=0.68, Conv2D=0.67) The logistic regression model showed superior performance and was chosen for the Body Response algorithm thanks to this performance and the lightweight and interpretable nature of its model weights.

These algorithms were all trained to predict stress labels from manually selected periods where the physiology showed stereotypical acute autonomic arousal responses (see the Stress Labels - TSST methods section for details), as the goal of this analysis was to create an algorithm that can identify the physiological dynamics associated with the onset of arousal responses, not simply baseline differences in stress vs no-stress periods. To demonstrate that these periods were not cherry-picked to boost performance, a logistic regression model was also fit to the labels generated based on the traditional TSST periods identified as stress and no-stress (see methods for details). This model performed better than the logistic regression model fit to the manually selected stress responses (ROC AUC=0.87, balanced accuracy=0.79, F1=0.70), but the features learned by this model captured fewer physiological dynamics. Thus, the logistic regression model trained on the manually selected stress responses was chosen for the Body Response algorithm and used in the subsequent modeling of the free-living data.

While optimal performance balancing false positives and false negatives is achieved with a probability threshold of 0.5, an alternate threshold may be desired when false positives are valued differentially from false negatives. Through extensive testing of subject preferences, a higher probability threshold of 0.72 was selected for the Body Response algorithm to reduce the number of false positives at the expense of false negatives. This resulted in a high specificity of 0.90 in the TSST training dataset, but a decreased F1 score of 0.54 and balanced accuracy of 0.66 (see Table 4 for all metrics).

|  |  | Model | | | |
|---|---|---|---|---|---|
|  |  | Conv2D | Random Forest | Linear SVM | Logistic Regression |
| Metric | ROC AUC | 0.73 | 0.73 | 0.79 | 0.79 |
|  | Accuracy | 0.68 | 0.69 | 0.72 | 0.73 |
|  | Bal. Accuracy | 0.67 | 0.68 | 0.73 | 0.73 |
|  | Sensitivity (Recall) | 0.62 | 0.66 | 0.74 | 0.75 |
|  | Specificity | 0.71 | 0.71 | 0.71 | 0.71 |
|  | PPV (Precision) | 0.59 | 0.60 | 0.64 | 0.64 |
|  | NPV | 0.74 | 0.75 | 0.80 | 0.81 |
|  | $F_1$-Score | 0.61 | 0.63 | 0.68 | 0.69 |
|  | FPR | 0.29 | 0.29 | 0.29 | 0.29 |

**Table 3.** Performance across all virtual lab TSST minutes of four different classification algorithms using either the non-linear aggregated features (Random Forest, SVM, Logistic Regression) or windows of input signals themselves (Conv2D). To enable fair comparison, a probability threshold was selected to pin FPR at 0.29 for all models. ROC AUC = Receiver Operating Characteristic Area Under the Curve, FPR = False Positive Rate, PPV = Positive Predictive Value, NPV = Negative Predictive Value.

|  |  | Model | | | |
|---|---|---|---|---|---|
|  |  | Conv2D | Random Forest | Linear SVM | Logistic Regression |
| Metric | ROC AUC | 0.60 | 0.73 | 0.79 | 0.79 |
|  | Accuracy | 0.66 | 0.67 | 0.71 | 0.71 |
|  | Bal. Accuracy | 0.60 | 0.62 | 0.67 | 0.66 |
|  | Sensitivity (Recall) | 0.30 | 0.34 | 0.43 | 0.42 |
|  | Specificity | 0.90 | 0.90 | 0.90 | 0.90 |
|  | PPV (Precision) | 0.67 | 0.69 | 0.75 | 0.75 |
|  | NPV | 0.65 | 0.67 | 0.70 | 0.70 |
|  | $F_1$-Score | 0.41 | 0.46 | 0.55 | 0.54 |
|  | FPR | 0.10 | 0.10 | 0.10 | 0.10 |

**Table 4**. Leave-one-subject-out training performance across 4 ML algorithms, holding FPR to 0.10 which corresponds to the Logistic Regression model's FPR when using the probability threshold selected for the production Body Response algorithm (0.72).

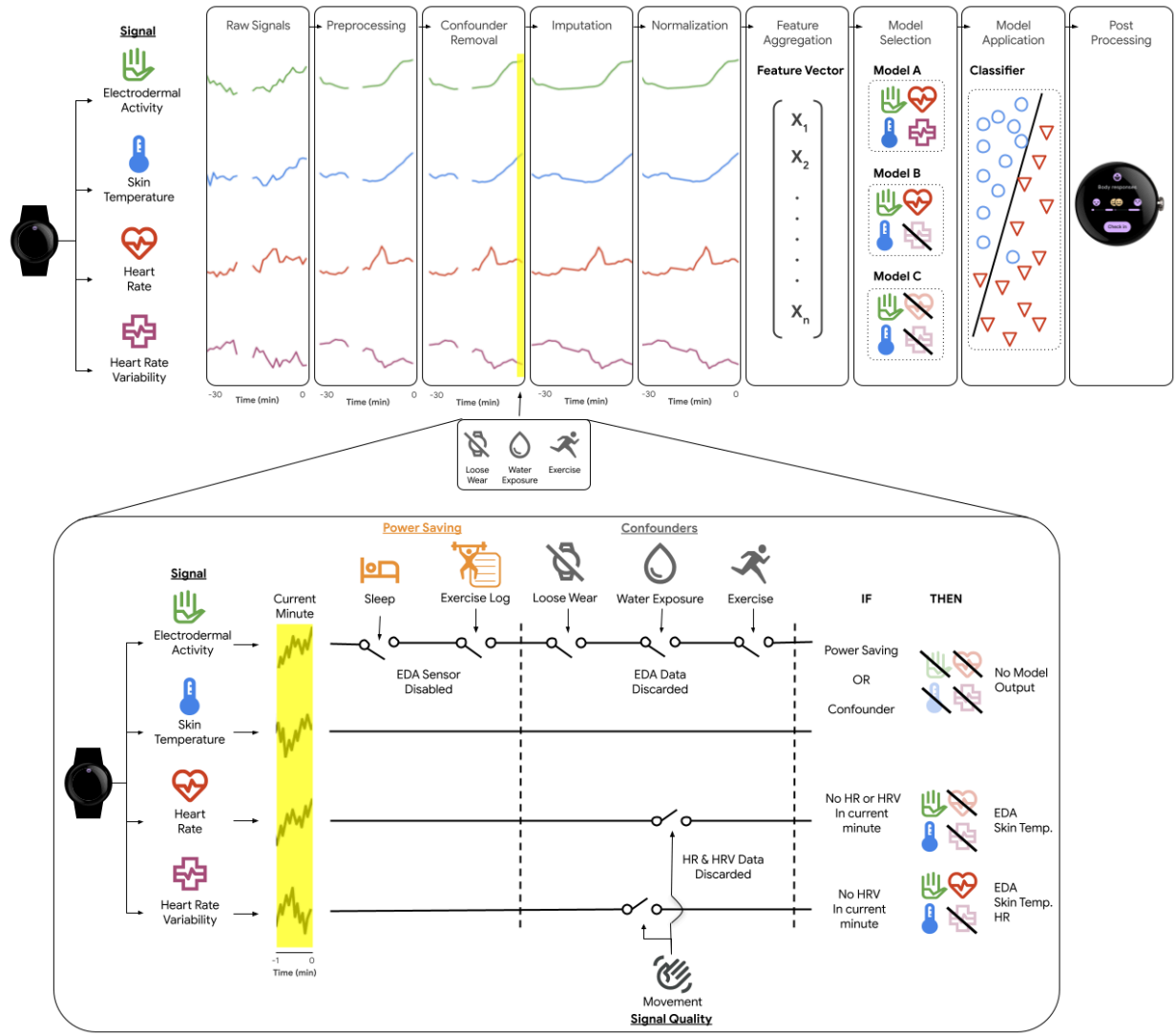

**Fig. 2.** Signal processing and stress classification pipeline. A high-level schematic of the steps involved in the Body Response algorithm's processing pipeline. The EDA, skin temperature, HR and HRV signals are initially filtered in a preprocessing step. Following this, confounded data is removed using information about sleep, exercise, loose wear and water exposure. The resulting, often partial, signals are imputed and normalized. Features are then extracted that form the input to the classifier.

*D. Free-Living*

A stress detection algorithm must be ecologically valid in order to provide utility beyond the lab, and so performance of the Body Response algorithm was validated on ecological momentary assessments (EMAs) of perceived stress during free-living. However, in free-living conditions there are many situations in which signals predictive of autonomic arousal can become confounded and no longer index arousal, such as exercise, water exposure or loose device wear (see Figure 2). Furthermore, different sensors can be affected in different ways by these confounders. For example, in 38.9% of minutes in our free-living data collect the PPG signals were noisy enough that it was not possible to reliably measure HRV. In 9.8% of minutes the HR signal was confounded by movement and HR measurements were not possible. Thus, to allow for the use of the largest ensemble of signals, while also providing maximal algorithmic availability, three LR models were fit and used in the Body Response algorithm, each subsequent model adding decreasingly available signals: 1) EDA + Temp, 2) EDA + Temp + HR, 3) EDA + TEMP + HR + HRV.

Comparison of LOSO training performance across the 3 LR models (see Table 5) shows each additional signal modality used incrementally improved performance. Balanced accuracy improved from 0.64 for the EDA + Temp model to 0.72 for the EDA + Temp +HR model, to 0.73 for the full EDA + Temp + HR + HRV model. This increase in performance from the addition of cardiovascular signals was larger for sensitivity (0.64 vs 0.73 vs 0.75) than for specificity (0.64 vs. 0.71 vs 0.71).

Table 6 shows the performance of the three models, plus the full Body Response algorithm, on the free-living data collection. Here, the higher probability threshold of 0.72 was used, resulting in higher specificity at the cost of sensitivity. Overall the three models performed somewhat similarly and accuracy was commensurate with the training dataset (balanced accuracy=0.61-0.69). As was seen in the training data, the EDA + Temp model was less sensitive (0.25) than the model including HR (0.40) or HR+HRV (0.37) but was more specific (0.97 vs 0.94 when adding HR or HR+HRV). The Body Response algorithm combines all 3 of these models, using the largest model whose sensor data is available. Combining all 3 models provides coverage over 100% of the minutes available to any model, and increases sensitivity (.47) at the expense of specificity (0.91).

The perceived stress EMAs used in the free-living dataset were sampled sparsely throughout the day, and even when they indicated stress were not guaranteed to align with the onset of any physiological arousal that may have occurred. To account for potential time alignment errors between stress EMAs and moments detected as stress by the Body Response algorithm, any positive stress EMA that came within +/- 10 minutes of a detected physiological stress response was deemed to be correctly identified (see methods for details). This adjustment means that chance performance in the free-living dataset may not be 50%. A permutation test was conducted, randomly shuffling detected stress responses throughout each subject-day, from which chance performance was estimated (Body Response Algorithm: balanced accuracy=63%, F1=0.36). Model performance of all 3 models, and the Body Response algorithm, were all significantly above chance (ps < 0.001 via 1000 sample one-tailed permutation test across all metrics).

|  |  | Logistic Regression Model | | |
|---|---|---|---|---|
|  |  | EDA+Temp | EDA+Temp+ HR | EDA+Temp+ HR+HRV |
| Metric | ROC AUC | 0.68 | 0.77 | 0.79 |
|  | Accuracy | 0.64 | 0.72 | 0.73 |
|  | Bal. Accuracy | 0.64 | 0.72 | 0.73 |
|  | Sensitivity (Recall) | 0.64 | 0.73 | 0.75 |
|  | Specificity | 0.64 | 0.71 | 0.71 |
|  | PPV (Precision) | 0.55 | 0.63 | 0.64 |
|  | NPV | 0.72 | 0.79 | 0.81 |
|  | $F_1$-Score | 0.59 | 0.67 | 0.69 |
|  | FPR | 0.36 | 0.29 | 0.29 |

**Table 5.** Leave-one-subject-out TSST training performance across all 3 Logistic Regression Models. Performance shown using a 0.5 probability threshold across models.

|  |  | Logistic Regression Model | | | |
|---|---|---|---|---|---|
|  |  | EDA+ Temp | EDA+ Temp+ HR | EDA+ Temp+ HR+ HRV | Body Response Algorithm (All 3) |
| Metric | Signal Availability* | 100.0% | 90.2% | 61.1% | 100.0% |
|  | Total Minutes | 39,675 | 35,780 | 24,271 | 39,675 |
|  | Accuracy | 0.87 | 0.86 | 0.86 | 0.85 |
|  | Bal. Accuracy | 0.61 | 0.67 | 0.66 | 0.69 |
|  | Sensitivity (Recall) | 0.25 | 0.41 | 0.37 | 0.47 |
|  | Specificity | 0.97 | 0.94 | 0.94 | 0.91 |
|  | PPV (Precision) | 0.58 | 0.53 | 0.53 | 0.47 |
|  | NPV | 0.89 | 0.91 | 0.90 | 0.91 |
|  | $F_1$-Score | 0.35 | 0.46 | 0.44 | 0.47 |

**Table 6.** Logistic Regression Model Performance on Perceived Stress Free-Living Data. Detected stress events that were within +/- 10 minutes of perceived stress labels were deemed correct.
NOTE: Signal Availability is calculated as a percentage of total available signal minutes, not total study time.

## V. DISCUSSION

In the present study we investigated the predictive power of physiological signals from wearable devices to identify moments of acute autonomic arousal in both a controlled and a real world setting. We described the development and validation of the Body Response algorithm, which identifies acute physiological stress responses via an ensemble of wearable sensor signals while also accounting for known stress confounders such as exercise, water exposure and loose device wear. Development of a real-world algorithm to detect a latent construct such as acute stress, one which has no agreed upon ground truth, requires varied testing domains that can address both criterion and ecological validity.

To demonstrate criterion validity, we trained and validated the Body Response algorithm to identify moments of confirmed physiological arousal induced via the "gold standard" Trier Social Stress Test (TSST). Inspection of the physiological signal dynamics induced by the TSST (Figure 1) revealed expected changes in HR, EDA and HRV as previously reported in the extensive TSST literature. These dynamics can only be captured by analysis of signals across a window of time, and so non-linear transformations of the sensor signals across 30-minute windows were calculated before being ensembled in ML models. Comparison across 4 commonly used ML models revealed slightly better performance in the two linear models relative to the 2 non-linear models, perhaps due to the increased bias of those models on features that had already been non-linearly aggregated. Based on minimal differences in the results of the 2 linear models, we selected logistic regression for inclusion in the Body Response algorithm, the most parsimonious and interpretable model.

The relationship between perceived and physiological stress responses are known to be modest [8], and so it is crucial to demonstrate that the Body Response algorithm not only has criterion validity in distinguishing physiological stress responses in controlled settings, but also ecological validity in distinguishing perceived stress in real world settings. To this end the Body Response algorithm was validated on an additional dataset during 1 week of free-living conditions. Real world conditions introduce challenges that differentially affect the signal quality of EDA, HR and HRV data streams, and thus how reliably available each is throughout the day. A model combining EDA and skin temperature was less sensitive to environmental and behavioral perturbations when compared to cardiovascular indices that are commonly used to index stress. To allow for maximal algorithmic on-time while also providing maximal performance at any given minute, three different ensemble models were used that combined an increasing number of sensor modalities. Performance classifying perceived stress in real world settings was good overall and comparable across our 3 models, but as models incorporated cardiovascular signals there was a slight increase in sensitivity and decrease in specificity. This increase in sensitivity for models including cardiovascular signals matched performance increases on the ground truth TSST training dataset across the 3 LR models trained. This increase in sensitivity at the expense of specificity matched our expectations that cardiovascular indices should be more sensitive and less specific, and was most pronounced when comparing performance on all available free-living minutes between the EDA + Temp model and the full Body Response algorithm (0.25 vs 0.47 sensitivity). Thus incorporating cardiovascular signals that do discretely shift between states of sympathetic and parasympathetic arousal in response to stress, alongside EDA signals that vary only in response to sympathetic arousal, results in a more sensitive model. Additionally, the reduced sensitivity observed with the EDA + Temp based models may in part be explained by the fact that a subset of the population are stabile and show less obvious skin conductance responses [21], especially when using dry electrodes.

The Body Response algorithm allows for near real-time identification of acute stress responses which, when delivered via notifications to wearable device users, affords the opportunity for various stress management interventions. When a stress event is detected, the Fitbit Sense 2 offers users the opportunity to reflect on how they're feeling, log their mood, and then further reflect or manage their stress via a guided breathing session, guided or free-form meditation, or a walk. Stress responses occur when a scenario is appraised as having high survival value [22], and so directing user attention to these scenarios can enable increased mindfulness which can be an effective stress management intervention [23], [24]. Through awareness, insight into potential negative beliefs or thoughts which resulted in the stress appraisal can be integrated and reframed, reducing future appraisal of similar situations as stressful [25]. Autonomic arousal is not only the result of negative appraisals, as positively valenced appraisal generates physiological changes similar to acute stress responses [26]. By offering both positive and negative emotions as mood logging options, we invite users to focus on positive events in their lives, potentially increasing user gratitude and reducing stress [27]. Additionally, well-timed brief interventions such as guided breathing [28], short walks [29], or just taking a break can help alleviate the negative symptoms of acute stress responses. These interventions illustrate what is possible when wearable devices can help bring awareness of autonomic arousal to individuals within a technological milieu with numerous other promising avenues for intervention. The design of the Body Response algorithm is consistent with guidelines that promote the safe and effective use of affective sensing [30].

*A. Limitations*

The study reported here does have some limitations that are important to consider when interpreting these findings. As with any study utilizing wrist-worn wearables as physiological sensors, there is a great deal of variation in how properly seated subjects affix their devices upon their wrists. Devices worn too loosely can lead to degraded signal quality across PPG-based cardiovascular signals [31]. These concerns are raised for electrodermal activity signals that require close skin contact given the need for a closed circuit across the two EDA electrodes [32]. Generally, these concerns more acutely affect free living conditions, where less experimenter intervention is possible, than controlled lab studies. While our study contained both a free living and controlled portion, the controlled TSST study was done virtually, over video conferencing software, due to shelter in place restrictions during the COVID-19 pandemic. While previous studies have demonstrated the validity of virtual TSST protocols [15], and the experimenters did their best to confirm proper wearable positioning over the video call, it is worth noting that some loose wear was still observed. Finally, while free living studies do confer more ecological validity than controlled studies, perceived stress EMA labels generally reflect perceived stress across many more minutes than the single point in time sampled, and so the labels do not perfectly capture minute to minute variation in subjective state. Additionally, any physiological responses that generated the perception of stress are not necessarily yoked in time to the EMA samples. To compensate for this we allowed a detected stress event from the Body Response algorithm to count as correct if it was within 10 minutes of the EMA label. This is a somewhat arbitrary cutoff, and so while we provided an estimate of its effects on chance performance, and demonstrated significance of our findings, it is worth noting that the ecological validity gained during free-living comes at the cost of label quality.

*B. Conclusion*

To conclude, these results demonstrate that autonomic arousal can be indexed and separated from environmental noise via ensemble models of wearable sensor signals. This was shown both within a controlled setting where stereotypical physiological responses were observed in response to known stressors, and in real world conditions where detected physiological responses were predictive of perceived stress. Inclusion of cardiovascular signals which index both sympathetic and parasympathetic arousal increased model sensitivity above models leveraging only sympathetically mediated EDA signals and skin temperature. Overall, this effort demonstrates the efficacy of leveraging wearable signals to identify acute stress responses in real world settings. It both demonstrates the promise of translating laboratory studies into real-world technologies that can help reduce human suffering and introduces a novel means by which to advance the study of physiological stress and psychological functioning embedded in real world contexts.

**Samy Abdel-Ghaffar**, is a Senior Research Scientist at Google. His research aims are to build technologies and do basic science that improve the mental and physical health of people. He received his PhD at the University of California, Berkeley in cognitive neuroscience in 2018 and bachelor's degree in computer science from the University of Southern California in 2001.

**Isaac Galatzer-Levy**, photograph and biography not available at the time of publication.

**Conor Heneghan** (Senior Member, IEEE) is a Senior Staff Research Scientist at Google. He completed his PhD at Columbia University, NY in 1995, and also holds a BEng from University College Dublin (1990). His research interests are in biomedical signal processing and analysis, and wearables.

**Xin Liu** (Member, IEEE), is a Senior Research Scientist at Google and a Research Affiliate at the University of Washington. His research lies in the intersection of machine learning and mobile computing. He has authored over 30 peer-reviewed papers in premium venues across machine learning, mobile and ubiquitous computing, and biomedical engineering. He received his Ph.D. in computer science from the University of Washington Seattle in 2023 and bachelor's with highest honors from the University of Massachusetts Amherst in 2018.

**Sarah Kernasovskiy**, photograph and biography not available at the time of publication.

**Brennan Garrett**, photograph and biography not available at the time of publication.

**Andrew Barakat**, photograph and biography not available at the time of publication.

**Daniel McDuff** (Member, IEEE), is a Staff Research Scientist at Google and an Affiliate Professor at the University of Washington. His research aims are to create technology that promotes the health and well-being of people and the planet. He completed his Ph.D. at the MIT Media Lab in 2014 and has a B.A. and M.S. from Cambridge University. Dr. McDuff has published over 150 peer-reviewed papers on affective computing, machine learning, human-computer interaction, and biomedical engineering. He was one of the pioneers of heart function measurements via visual imaging of the face.